\documentclass[prl,english,preprintnumbers,nofootinbib,twocolumn,superscriptaddress]{revtex4-1}
\usepackage{graphicx}
\usepackage{float}% floating figures at the exact position
\usepackage{mathtools}
\usepackage{amssymb}
\usepackage{accents}
\usepackage{verbatim}%for multiple line commenting
\usepackage{hyperref}
\usepackage{xcolor}
\hypersetup{
    colorlinks,
    linkcolor={red!50!black},
    citecolor={blue!50!black},
    urlcolor={blue!80!black}
}

\newcommand{\T} {{\scriptscriptstyle T}}
\newcommand{\ubar}[1]{\underaccent{\bar}{#1}}
\newcommand{\gf} {G_{\!f}}
\newcommand{\gfp} {G_{\mathclap{f}}^p}
\newcommand{\gfc} {\tilde{G}_{\mathclap{f}}^p}
\newcommand{\hfp} {H_{\mathclap{f}}^p}
\newcommand{\hfc} {\tilde{H}_{\mathclap{f}}^p}
\newcommand{\gfour} {G_{\mathclap{f}}^4}
\newcommand{\gfourc} {\tilde{G}_{\mathclap{f}}^4}

\newcommand {\cy}{Z}
\newcommand{\rs} {\boldsymbol{1}}
\newcommand{\rsp} {\boldsymbol{\ubar{1}}}
\newcommand{\rsa} {\acute{\boldsymbol{\ubar{1}}}}
\newcommand{\rsg} {\grave{\boldsymbol{\ubar{1}}}}
\newcommand{\rd} {\boldsymbol{2}}
\newcommand{\rda} {\acute{\boldsymbol{2}}}
\newcommand{\rdg} {\grave{\boldsymbol{2}}}
\newcommand{\rt} {\boldsymbol{3}}
\newcommand{\rtp} {\boldsymbol{\ubar{3}}}
\newcommand {\phic}{\phi^{\mathclap{\,\,\hskip -.05em\circ}\,}}
\newcommand{\phia} {\acute{\phi}}
\newcommand{\phih} {\hat{\phi}}
\newcommand{\phig} {\grave{\phi}}
\newcommand{\phisa} {\acute{\phi}_s}
\newcommand{\phish} {\hat{\phi}_s}
\newcommand{\phisg} {\grave{\phi}_s}
\newcommand{\phiac} {\acute{\phi}^{\circ\!}}
\newcommand{\phihc} {\hat{\phi}^{\circ\!}}
\newcommand{\phigc} {\grave{\phi}^{\circ\!}}
\newcommand{\phisac} {\acute{\phi}_s^{\circ\!}}
\newcommand{\phishc} {\hat{\phi}_s^{\circ\!}}
\newcommand{\phisgc} {\grave{\phi}_s^{\circ\!}}

\begin{document}
	
\title{Symmetries of stationary points of the $G$-invariant potential and \\the framework of the auxiliary group}
\author{R.~Krishnan}
\email{krishnan.rama@saha.ac.in}
\homepage{\\https://orcid.org/0000-0002-0707-3267}
\affiliation{Saha Institute of Nuclear Physics, 1/AF Bidhannagar, Kolkata 700064, India}  
\begin{abstract}
We classify the constraints on a stationary point of the potential invariant under a finite group into intrinsic and extrinsic based on whether they are independent of the coefficients in the potential or not. We find that the symmetry group of a set of stationary points can be larger than that of the potential and the stabilizer under this group generates intrinsic constraints. By applying these findings in the framework of the auxiliary group, we show that the constraints that can only be obtained extrinsically in an elementary theory can be generated intrinsically in an effective theory\footnote{A video presentation of this paper is available \href{https://www.youtube.com/playlist?list=PLjRYJtC1E1HdCLC8myTbioUMLZpUmaXBL}{here}.}.
 
\end{abstract}
\maketitle

\centerline{\bf \hspace{2mm}I.\hspace{2mm}Introduction}
\vspace{1.5mm}
Discrete symmetries implemented using finite groups have been used extensively  \cite{1510.02091, 1711.10806} in particle physics to study flavor. Scalar fields called flavons, which transform as multiplets under a finite group, are often introduced. Their potential is constructed, and one of the stationary points (SPs) of the potential is chosen as their vacuum expectation value (VEV), leading to spontaneous symmetry breaking (SSB). The VEV plays a central role in determining the structure of the fermion mass matrices. SSB of finite groups appears in the context of dark matter fields \cite{1007.0871, 1008.4777} and inflatons \cite{hep-ph/0001333, 1407.6017, 1412.8537, 1808.09601} in cosmology and various fields \cite{GAETA2006322, MICHEL200111} in solid-state physics also. Therefore, understanding the symmetries of SPs of the potential is crucial in a wide variety of settings.

\vspace{5mm}
\centerline{\bf \hspace{2mm}II.\hspace{2mm}Constraints on SPs}
\vspace{1.5mm}
Consider a field $\phi$ transforming as a multiplet of a $d$-dimensional orthogonal real representation of a finite group $\gf$. The invariants of the group action constructed using $\phi$ constitute the potential. Let the number of linearly independent invariants up to order $p$ be $N$. The most general potential of order $p$ is given by
\begin{equation}\label{eq:generalpot}
\mathcal{V}^p = c_\alpha \mathcal{I}_\alpha,
\end{equation}
where $\mathcal{I}_\alpha$ are the invariants and $c_\alpha$ are the corresponding real coefficients. Invariants of mass dimension up to four are renormalizable. Since higher-order invariants are suppressed, we have $p\geq4$. Let $\gfp$ be the largest group under which $\phi$ transforms as an orthogonal real representation such that the potential $\mathcal{V}^p$ remains invariant. In general, we have $\gfp \geq \gf$. In our analysis, we assume that $p$ is sufficiently high so that $\gfp$ is a finite group. 

The set of coefficients $c=(c_1,..,c_N)$ and the field $\phi = (\phi_1,..,\phi_d)$ form the topological spaces $\mathbb{R}^N$ and $\mathbb{R}^d$, respectively. Let $\mathcal{C}$ be an open subset of $\mathbb{R}^N$ with $c \in \mathcal{C}$ such that we can define a map
\begin{equation}\label{eq:map}
\phic:\mathcal{C}\rightarrow \mathbb{R}^d:\quad c_\alpha\,\partial_i \mathcal{I}_\alpha\big|_{\phi=\phic(c)} =0,
\end{equation}
where $\partial_i$ is the derivative with respect to the component $\phi_i$. Eq.(\ref{eq:map}) implies that $\phic(c)$ is a SP of $\mathcal{V}^p$ constructed with the coefficients $c$.
Moreover, we require the SP to be nondegenerate \cite{Bott}, i.e.~the Hessian matrix $\quad c_\alpha\,\partial_j \partial_i \mathcal{I}_\alpha\big|_{\phi=\phic(c)}$ is assumed to be nonsingular. This requirement ensures that there exists an open subset in $\mathbb{R}^d$ such that $\phic(c)$ is the only SP in it for a given $c$, and as a result the map $\phic(c)$ is uniquely defined.

Through SSB, $\phic(c)$ can become the VEV of $\phi$. In flavor models, constraints on $\phic(c)$ result in constraints on the mass matrices leading to interesting predictions about the observables. Since these constraints are at the heart of such models, let us study them in detail. If
\begin{equation}\label{eq:intrinsic}
\mathcal{F}(\phic(c)) = 0 \quad \forall c \in \mathcal{C},
\end{equation}
where $\mathcal{F}$ is a function that does not explicitly depend on $c$, we call (\ref{eq:intrinsic}) an intrinsic constraint. An intrinsic constraint remains valid irrespective of any change in the coefficients (within $\mathcal{C}$ where the map $\phic(c)$ is uniquely defined). Its genesis is totally in the group theoretical structure of the system. 

Suppose we impose $(N-n)$ constraints of the form $f_i(c)=0$, where $f_i$ are functions with $i=1,..,N-n$, on the coefficients. Let $\mathcal{C}_n$ be the $n$-dimensional subspace of $\mathcal{C}$ that satisfies these constraints. Let $\phic(c)_n$ be the restriction of $\phic(c)$ to the domain $\mathcal{C}_n$. If (\ref{eq:intrinsic}) is not valid, i.e.,~$\exists c \in \mathcal{C} : \mathcal{F}(\phic(c))\neq 0$, and if
\begin{equation}\label{eq:extrinsic}
\mathcal{F}(\phic(c)_n) = 0 \quad \forall c \in \mathcal{C}_n,
\end{equation}
then we call (\ref{eq:extrinsic}) an extrinsic constraint. Rather than being the result of the group theoretical structure, an extrinsic constraint is an artifact of the restriction. As an example, consider the model described in \cite{hep-ph/0504165} which consists of two triplets and a singlet of $A_4$, i.e.,~$\phi=((\varphi_1,\varphi_2,\varphi_3), (\varphi'_1,\varphi'_2,\varphi'_3),\xi)$. The authors set certain cross-couplings between the two triplets to vanish, i.e.,~they imposed constraints of the form $f_i(c)=0$, so as to obtain the desired SP. This SP, which is nothing but the restriction of $\phic(c)$ to the lower dimensional domain $\mathcal{C}_n$, was given by $\phic(c)_n=((v(c),v(c),v(c)), (v'(c),0,0), u(c))$ where $v(c), v'(c)$ and $u(c)$ were defined in $\mathcal{C}_n$. This corresponds to four independent extrinsic constraints: $\varphi^{\mathrlap{\circ}}_{1\!}(c)_n-\varphi^{\mathrlap{\circ}}_{2\!}(c)_n=0$, $\varphi^{\mathrlap{\circ}}_{1\!}(c)_n-\varphi^{\mathrlap{\circ}}_{3\!}(c)_n=0$, $\varphi^{\mathrlap{\hskip -.03em \prime\hskip -.03em\circ}}_{2}(c)_n=0$ and $\varphi^{\mathrlap{\hskip -.03em \prime\hskip -.03em\circ}}_{3}(c)_n=0$. 

Neutrino flavor models widely impose constraints on the coefficients to obtain extrinsic constraints. They include the ones like \cite{hep-ph/0504165, hep-ph/0511133, 0811.0345, 2003.00506} which use them to decouple the SPs of irreducible multiplets (irreps) that are associated with different fermion mass terms and others like \cite{hep-ph/0610250, 0912.1344, 1801.10197} which use them to obtain constraints within the SP of a single irrep. Even though some of these models give arguments (which are perhaps not very convincing) to justify their imposition, we advocate the group theoretically elegant approach in which all constraints are obtained intrinsically. However, such an approach will be quite difficult because the experimental data points towards a complete breaking of $\gf$ while intrinsic constraints have always been found to be associated with the unbroken part (the stabilizer) of $\gfp$.

\vspace{5mm}
\centerline{\bf \hspace{2mm}III.\hspace{2mm}Intrinsic constraints generated from the}
\centerline{\bf stabilizer under $\gfp$}
\vspace{1.5mm}
The stabilizer \cite{MICHEL200111} (often referred to as the residual symmetry group) of $\phi$ under $\gfp$ is given by
\begin{equation}\label{eq:hH}
\hfp=\{h\in \gfp : h_{ij} \phi_j = \phi_i \}.
\end{equation}
In this section, we will show that the stabilizer of a SP generates intrinsic constraints. We define
\begin{equation}\label{eq:lambda}
\Lambda_{i\alpha}\hskip 1.1em\mathllap{(c)} = \left(\phic_{\mathclap{\,i}}(c+\delta c)-\phic_{\mathclap{\,i}}(c)\right) /\delta c_\alpha,
\end{equation}
where $\delta c$ denotes a change of $c$ with $(c+\delta c) \in \mathcal{C}$. Applying (\ref{eq:map}) at $(c+\delta c)$ gives $(c_\alpha+\delta c_\alpha)\partial_i \mathcal{I}_\alpha\big|_{\phi=\phic(c+\delta c)}=0$. Taylor expanding this equation around $c$ and using (\ref{eq:lambda}), we obtain
\begin{equation}\label{eq:taylor}
\begin{split}
&(c_\alpha + \delta c_\alpha)\left(\partial_i \mathcal{I}_\alpha\big|_{\phi=\phic(c)}+\partial_j \partial_i \mathcal{I}_\alpha\big|_{\phi=\phic(c)} \Lambda_{j\beta}\hskip 1.1em\mathllap{(c)}\delta c_\beta\right. \\
&\quad\quad\quad\quad\quad\quad\quad\quad\quad\quad\quad\quad\quad\quad\left.  \vphantom{\big|_\phi}+ \mathcal{O}(\delta c^2)\right)=0.
\end{split}
\end{equation}
A minimal set of points that remains invariant under the action of a group $G$ is called a $G$-orbit. SPs always form orbits with respect to the symmetry group of the potential \cite{MICHEL200111}, i.e.,~the SPs of $\mathcal{V}^p$ exist in $\gfp$-orbits,
\begin{equation}\label{eq:gkinvariant}
(c_ \alpha+\delta c_ \alpha)\,\partial_i \mathcal{I}_\alpha\big|_{\phi=g\phic(c+\delta c)} =0 \quad \forall \, g \in \gfp.
\end{equation}
Taylor expanding (\ref{eq:gkinvariant}) and using (\ref{eq:lambda}), we obtain
\begin{align}\label{eq:taylor2}
		&(c_ \alpha+\delta c_ \alpha)\left(\partial_i \mathcal{I}_\alpha\big|_{\phi=g\phic(c)}+\partial_j \partial_i \mathcal{I}_\alpha\big|_{\phi=g\phic(c)} g_{jk}\Lambda_{k\beta}\hskip 1.1em\mathllap{(c)}\delta c_\beta \right. \notag\\
		&\quad\quad\quad\quad\quad\quad\quad\quad\quad\quad+ \left. \vphantom{\big|_\phi}\mathcal{O}(\delta c^2)\right)=0 \quad  \forall \, g \in \gfp.
\end{align}
Comparing the terms linear in $\delta c$ in (\ref{eq:taylor}) and (\ref{eq:taylor2}), we obtain
\begin{equation}\label{eq:linear}
\begin{split}
&\partial_i \mathcal{I}_\beta\big|_{\phi=\phic(c)}+c_\alpha\partial_j \partial_i \mathcal{I}_\alpha\big|_{\phi=\phic(c)} \Lambda_{j\beta}\hskip 1.1em\mathllap{(c)}\\
&\quad \quad=\partial_i \mathcal{I}_\beta\big|_{\phi=g\phic(c)}+c_\alpha\partial_j \partial_i \mathcal{I}_\alpha\big|_{\phi=g\phic(c)} g_{jk}\Lambda_{k\beta}\hskip 1.1em\mathllap{(c)}
\end{split}
\end{equation}
Let us assume that $\hfp$ is the stabilizer of $\phic(c)$ under $\gfp$ for a specific value of $c$, i.e.,~$\exists c \in \mathcal{C}: h\phic(c)=\phic(c) \,\,\forall h \in \hfp$. Replacing $g$ with $h$ in (\ref{eq:linear}) results in
\begin{equation}\label{eq:productzero}
c_\alpha\partial_j \partial_i \mathcal{I}_\alpha\big|_{\phi=\phic(c)} \left( \vphantom{\big|_\phi}\Lambda_{j\beta}\hskip 1.1em\mathllap{(c)}-h_{jk} \Lambda_{k\beta}\hskip 1.1em\mathllap{(c)}\right) =0.
\end{equation}
Since the Hessian matrix $c_\alpha\partial_j \partial_i \mathcal{I}_\alpha\big|_{\phi=\phic(c)}$ is assumed to be nonsingular, we obtain
\begin{equation}\label{eq:hlambda}
h_{jk} \Lambda_{k\beta}\hskip 1.1em\mathllap{(c)} = \Lambda_{j\beta}\hskip 1.1em\mathllap{(c)} \quad \forall \, h \in \hfp.
\end{equation}
The action of $h$ on $\phic(c+\delta c)$ is given by
\begin{equation}\label{eq:proof}
h_{ij} \phic_{\mathclap{\,j}}(c+\delta c) = h_{ij} (\phic_{\mathclap{\,j}}(c)+\Lambda_{j\beta}\hskip 1.1em\mathllap{(c)} \delta c_\beta) = \phic_{\mathclap{\,i}}(c+\delta c),
\end{equation}
i.e.,~if $\hfp$ is the stabilizer of $\phic(c)$ for a specific value of $c$, it will be the stabilizer of $\phic(c+\delta c)$ for all $(c+\delta c) \in \mathcal{C}$. This results in intrinsic constraints of the form
\begin{equation}
(h-I)_{ij}\phic_{\mathclap{\,j}}(c)=0 \,\, \forall h \in \hfp,\, c \in \mathcal{C}
\end{equation}
where $I$ is the identity.

\vspace{5mm}
\centerline{\bf \hspace{2mm}IV.\hspace{2mm}Symmetry group of a set of SPs}
\vspace{1.5mm}
SPs form $\gfp$-orbits, i.e.,~$c_\alpha\,\partial_i \mathcal{I}_\alpha\big|_{\phi=g\phic(c)} =0 \quad \forall g \in \gfp, c \in \mathcal{C}$. Let us examine whether a union of such $\gfp$-orbits can have a symmetry group larger than $\gfp$ for all $c \in \mathcal{C}$. We define
\begin{equation}\label{eq:quasi}
\gfc=\{g : \,c_\alpha\,\partial_i \mathcal{I}_\alpha\big|_{\phi=g\phic(c)} =0 \quad \forall c \in \mathcal{C}\}.
\end{equation}
Given a SP $\phic(c)$, (\ref{eq:quasi}) implies the existence of a set of SPs $g\phic(c)$ with $g\in\gfc$ for all $c \in \mathcal{C}$. We expect $\gfc\geq\gfp$. Let us examine a scenario where $\gfc>\gfp.$\footnote{$\gfc$ being larger than $\gfp$ may not remain valid if we increase the order $p$. However, higher-order nonrenormalizable terms are suppressed. If we obtain a nondegenerate SP at a lower order, higher-order terms cause only small perturbations to it.}

We rewrite (\ref{eq:generalpot}) as
\begin{equation}\label{eq:pot2}
\mathcal{V}^p= \sum\nolimits_{\alpha=1}^n c_\alpha \mathcal{I}_\alpha + \sum \nolimits_{\alpha=n+1}^{N} c_\alpha \mathcal{X}_\alpha^\T\mathcal{Y}_\alpha,
\end{equation}
with the following three assumptions: i) The first summation consists of $n$ $\gfp$-invariants. We assume that they remain invariant under the larger group $\gfc$ also. ii) The second summation involves the rest of the $(N-n)$ $\gfp$-invariants, each of which is expressed as the dot product $\mathcal{X}_\alpha^\T\mathcal{Y}_\alpha$ where $\mathcal{X}_\alpha$ and $\mathcal{Y}_\alpha$ are multiplets of the same representation of $\gfp$. We assume that $\mathcal{X}_\alpha$ and $\mathcal{Y}_\alpha$ are multiplets of different representations of $\gfc$. Hence, $\mathcal{X}_\alpha^\T\mathcal{Y}_\alpha$ is not a $\gfc$-invariant. iii) Let $\mathcal{C}_n$ be the $n$-dimensional subspace of $\mathcal{C}$ obtained by imposing the constraints $c_i=0$ with $i=n+1,..,N$. Let $\mathcal{V}^p_n$ and $\phic(c)_n$ be the restrictions of $\mathcal{V}^p$ and $\phic(c)$ respectively to $\mathcal{C}_n$. We assume that
\begin{equation}\label{eq:assume3}
\mathcal{X}_\alpha\big|_{\phi=\phic(c)_n}=0,\,\,\, \mathcal{Y}_\alpha\big|_{\phi=\phic(c)_n}=0\quad\forall c \in \mathcal{C}_n. 
\end{equation}

Since $\mathcal{X}_\alpha$ and $\mathcal{Y}_\alpha$ are multiplets under $\gfc$, their vanishing at $\phi=\phic(c)_n$ implies
\begin{equation}\label{eq:part2}
	\partial_i(\mathcal{X}_\alpha^\T\mathcal{Y}_\alpha)\big|_{\phi=g\phic(c)_n}=0 \quad \forall \,g \in \gfc, \,c \in \mathcal{C}_n.
\end{equation}
Since $g\phic(c)_n\,\forall g \in \gfc, c \in \mathcal{C}_n$ are the SPs of $\mathcal{V}^p_n$, and given (\ref{eq:part2}), we obtain
\begin{equation}\label{eq:part1}
\begin{split}
&\partial_i\left(\sum\nolimits_{\alpha=1}^n c_\alpha \mathcal{I}_\alpha + \sum\nolimits_{\alpha=n+1}^N c_\alpha \mathcal{X}_\alpha^\T\mathcal{Y}_\alpha\right)\big|_{\phi=g\phic(c)_n}\\
&\quad\quad\quad\quad\quad\quad\quad\quad\quad\quad\quad\quad =0 \quad \forall g \in \gfc, c \in \mathcal{C}_n,
\end{split}
\end{equation}
i.e.,~$g\phic(c)_n \forall g \in \gfc, c \in \mathcal{C}_n$ are the SPs of not only $\mathcal{V}^p_n$ but also $\mathcal{V}^p$. The last $(N-n)$ coefficients of $\mathcal{V}^p$ do not play any role in determining these SPs of $\mathcal{V}^p$. Therefore, we obtain $\phic(c)=\phic(c)_n$ with $\mathcal{C}=\mathcal{C}_n\times\mathbb{R}^{N-n}$. 

\vspace{5mm}
\centerline{\bf \hspace{2mm}V.\hspace{2mm}Intrinsic constraints generated from the}
\centerline{\bf stabilizer under  $\gfc$}
\vspace{1.5mm}
Let $\hfc$ be a subgroup of $\gfc$ such that it is the stabilizer of $\phic(c)$ under $\gfc$ for a specific value of $c \in \mathcal{C}$. Following the steps similar to those in the previous section, we can obtain equations similar to (\ref{eq:gkinvariant})-(\ref{eq:proof}) in which $\gfp$ and $\hfp$ are replaced with $\gfc$ and $\hfc$, respectively, leading to the conclusion that $\hfc$ is the stabilizer of $\phic(c)$ for all $c \in \mathcal{C}$. As a result, we obtain intrinsic constraints of the form $(h-I)_{ij}\phic_{\mathclap{\,j}}(c)=0\, \forall h \in \hfc, c \in \mathcal{C}$. Thus we conclude that the stabilizer under not only $\gfp$ but also $\gfc$ generates intrinsic constraints. We now demonstrate these concepts with the help of an example. 

{\renewcommand{\arraystretch}{1.5}
	\begin{table}[]
		\begin{center}
			\begin{tabular}{|c| c c c|}
				\hline
				&$\phia,\phisa$&$\phih,\phish$&$\phig,\phisg$\\
				\hline
				$S_4\!\times\!\acute{D}_6\!\times\! \grave{D}_6$&$\rt\!\times\!\rda\!\times\!\rs,\rs$&$\rs\!\times\!\rda\!\times\!\rdg,\rs$&$\rt\!\times\!\rs\!\times\!\rdg,\rs$\\
				$\cy_2\!\times\!\acute{\cy}_2\!\times\! \grave{\cy}_2$&$1\!\times\!-1\!\times\!1$&$-1\!\times\!-1\!\times\!-1$&$1\!\times\!1\!\times\!-1$\\
				\hline
			\end{tabular}
			\caption{The action of $\gf =S_4\!\times\!\cy_2\!\times\!\acute{D}_6\!\times\!\acute{\cy}_2\!\times\!\grave{D}_6\!\times\!\grave{\cy}_2$ on the irreps $\phia$, $\phisa$, $\phih$, $\phish$, $\phig$ and $\phisg$.}
			\label{tab:flavorcontent}
		\end{center}
	\end{table}
}

\vspace{5mm}
\centerline{\bf VI.\hspace{2mm}Example} 
\vspace{1.5mm}
The finite group used in this example consists of the symmetric group $S_4$ \cite{Pakvasa:1978tx, hep-ph/9403201, hep-ph/0508231, 2003.00506}, two dihedral groups of order six (named $\acute{D}_6$, $\grave{D}_6$) and three cyclic groups of order two (named $\cy_2$, $\acute{\cy}_2$, $\grave{\cy}_2$). We denote the irreps of $S_4$ by $\rs$, $\rsp$, $\rd$, $\rt$, $\rtp$; $\acute{D}_6$ by $\rs$, $\rda$, $\rsa$ and $\grave{D}_6$ by $\rs$, $\rdg$, $\rsg$. As a convenient basis, we use
\begin{equation*}
 S=\left(\begin{matrix} 1 & 0 &0\\
0 & -1 & 0\\
0 & 0 & -1
\end{matrix}\right)\!, \,T=\left(\begin{matrix} 0 & 1 & 0\\
0 & 0 & 1\\
1 & 0 & 0
\end{matrix}\right)\!, \,U=\left(\begin{matrix} 1 & 0 & 0\\
0 & 0 & 1\\
0 & 1 & 0
\end{matrix}\right)
\end{equation*}
and 
\begin{equation*}
s=\left(\begin{matrix} 1 & 0\\
0 & 1
\end{matrix}\right)\!, \, t= \left(\begin{matrix} -\frac{1}{2} & -\frac{\sqrt{3}}{2}\\
\frac{\sqrt{3}}{2} & -\frac{1}{2} 
\end{matrix}\right)\!, \, u=\left(\begin{matrix} -\frac{1}{2} & \frac{\sqrt{3}}{2}\\
\frac{\sqrt{3}}{2} & \frac{1}{2}
\end{matrix}\right)
\end{equation*}
to generate $\rt$ and $\rd$, $\rda$, $\rdg$, respectively. In our analysis, we utilize the tensor product expansions $\rt\times \rt = \rs+\rd+\rt+\rtp$, $\rda \times \rda=\rs+\rda+\rsa$ and $\rdg \times \rdg=\rs+\rdg+\rsg$ where $\rtp$, $\rsa$ and $\rsg$ are antisymmetric products while the rest are symmetric.

We propose a scalar field $\phi=(\phia,\phisa,\phih,\phish,\phig,\phisg)$, Table~\ref{tab:flavorcontent}, which transforms under our flavor symmetry group, $\gf =S_4\!\times\!\cy_2\!\times\!\acute{D}_6\!\times\!\acute{\cy}_2\!\times\!\grave{D}_6\!\times\!\grave{\cy}_2$. Since $\phia$, $\phih$ and $\phig$ transform as $\rt\!\times\!\rda$, $\rda\!\times\!\rdg$ and $\rdg\!\times\!\rt$ under $S_4\!\times\!\acute{D}_6$, $\acute{D}_6\!\times\!\grave{D}_6$ and $\grave{D}_6\!\times\!S_4$, we express them as $3\!\times\!2$, $2\!\times\!2$ and $2\!\times\!3$ real matrices, respectively. By taking their tensor products with themselves, we construct the following quadratic multiplets. Using $\phia$, we obtain
\begin{align*}
	|\phia|^2&= \text{Tr}[\phia^\T\phia], \quad \quad \phia^2_{\rd\rs} = \left(\text{Tr}[\phia^\T\lambda_8\phia],\text{Tr}[\phia^\T\lambda_3\phia]\right)^{\!\T}\!\!,\\
	\phia^2_{\rt\rs}&= \left(\text{Tr}[\phia^\T\lambda_6\phia], \text{Tr}[\phia^\T\lambda_4\phia], \text{Tr}[\phia^\T\lambda_1\phia]\right)^{\!\T}\!\!\!,\\ \phia^2_{\rs\rda} &= \left(\text{Tr}[\phia^\T\phia\sigma_3],-\text{Tr}[\phia^\T\phia\sigma_1]\right)\!,\\
	\phia^2_{\rd\rda}&=\left(\begin{matrix}\text{Tr}[\phia^\T\lambda_8\phia\sigma_3] & -\text{Tr}[\phia^\T\lambda_8\phia\sigma_1]\!\!\!,\\
		\text{Tr}[\phia^\T\lambda_3\phia\sigma_3] & -\text{Tr}[\phia^\T\lambda_3\phia\sigma_1]
	\end{matrix}\right),\\
	\phia^2_{\rt\rda}&=\left(\begin{matrix}\text{Tr}[\phia^\T\lambda_6\phia\sigma_3] & \text{Tr}[\phia^\T\lambda_4\phia\sigma_3] & \text{Tr}[\phia^\T\lambda_1\phia\sigma_3]\\
		-\text{Tr}[\phia^\T\lambda_6\phia\sigma_1] & -\text{Tr}[\phia^\T\lambda_4\phia\sigma_1] & -\text{Tr}[\phia^\T\lambda_1\phia\sigma_1]
	\end{matrix}\right)^{\!\!\T}\!\!\!,\\
	\phia^2_{\rtp\rsa} &= \left(\text{Tr}[\phia^T\lambda_7\phia\sigma_2], -\text{Tr}[\phia^\T\lambda_5\phia\sigma_2], \text{Tr}[\phia^\T\lambda_2\phia\sigma_2]\right)^{\!\T}\!\!,
\end{align*}
where $\sigma_i$ and $\lambda_i$ are the Pauli matrices and the Gell-Mann matrices, respectively. The naming of these multiplets is self-explanatory, e.g.,~$\phia^2_{\rd\rda}$ is the multiplet quadratic in $\phia$ and transforming as $\rd$ and $\rda$ under $S_4$ and $\acute{D}_6$, respectively. We obtain similar quadratic multiplets using $\phig$ also: $|\phig|^2$, $\phig^2_{\rs\rd}$, $\phig^2_{\rs\rt}$, $\phig^2_{\rdg\rs}$, $\phig^2_{\rdg\rd}$,  $\phig^2_{\rdg\rt}$ and $\phih^2_{\rsg\rtp}$. The quadratic multiplets obtained using $\phih$ are
\begin{align*}
	|\phih|^2&= \text{Tr}[\phih^\T\phih], &\phih^2_{\rs\rdg}&=\left(\text{Tr}[\phih^\T\phih\sigma_3],-\text{Tr}[\phih^\T\phih\sigma_1]\right),\\
	\phih^2_{\rsa \rsg} &= \text{Tr}[\phih^\T\sigma_2\phih\sigma_2], &\phih^2_{\rda\rs}&=\left(\text{Tr}[\phih^\T\sigma_3\phih],-\text{Tr}[\phih^\T\sigma_1\phih]\right)^{\!\T}\!\!,
\end{align*}
\vspace{-0.71cm}
\begin{equation*}\label{eq:phimat2}
	\phih^2_{\rda\rdg}=\left(\begin{matrix}\text{Tr}[\phih^\T\sigma_3\phih\sigma_3] & -\text{Tr}[\phih^\T\sigma_3\phih\sigma_1]\\
		-\text{Tr}[\phih^\T\sigma_1\phih\sigma_3] & \text{Tr}[\phih^\T\sigma_1\phih\sigma_1]
	\end{matrix}\right).
\end{equation*} 

Now, we construct the invariants. The quadratic invariants are $|\phia|^2$, $|\phisa|^2$, $|\phih|^2$, $|\phish|^2$, $|\phig|^2$ and $|\phisg|^2$. By taking the product of any two of these, we obtain $21$ quartic invariants. We also have another 12 quartic invariants: $(\phia^2_{\rd\rs})^\T\phia^2_{\rd\rs}$, $(\phia^2_{\rt\rs})^\T\phia^2_{\rt\rs}$, $\phisa\text{Tr}[\phia^\T\phia^2_{\rt\rda}]$, $\phig^2_{\rs\rd}(\phig^2_{\rs\rd})^\T$, $\phig^2_{\rs\rt}(\phig^2_{\rs\rt})^\T$, $\phisg\text{Tr}[\phig^\T\phig^2_{\rdg\rt}]$, $(\phih^2_{\rsa\rsg})^2$,  $\phish\text{Tr}[\phih^\T\phih^2_{\rda\rdg}]$,  $\phig^2_{\rs\rd}\phia^2_{\rd\rs}$, $\phig^2_{\rs\rt}\phia^2_{\rt\rs}$, $\phia^2_{\rs\rda}\phih^2_{\rda\rs}$ and $\phih^2_{\rs\rdg}\phig^2_{\rdg\rs}$. These 39 invariants along with 39 coefficients constitute the most general potential of order $4$, i.e.,~for $p=4$, we construct $\mathcal{V}^4$ with $N=39$ terms. Note that in this example, we have $\gfour=\gf$.

Let us replace $\acute{D}_6$ in $\gfour$ with two different dihedral groups $\acute{D}^a_6$ and $\acute{D}^b_6$ acting on the rhs of $\phia$ and the lhs of $\phih$, respectively. Similarly, we replace $\grave{D}_6$ with $\grave{D}^a_6$ and $\grave{D}^b_6$ acting on the lhs of $\phig$ and the rhs of $\phih$, respectively. In anticipation of the things to come, let us name the resulting group $\gfourc$, i.e.,~$\gfourc=S_4\!\times\!\cy_2\!\times\!\acute{D}^a_6\!\times\!\acute{D}^b_6\!\times\!\acute{\cy}_2\!\times\!\grave{D}^a_6\!\times\!\grave{D}^b_6\!\times\!\grave{\cy}_2$. In relation to $\gfourc$, let us verify the three conditions listed in the previous section. 

(i) Every term except $\phia^2_{\rs\rda}\phih^2_{\rda\rs}$ and $\phih^2_{\rs\rdg}\phig^2_{\rdg\rs}$ is invariant under $\gfourc$, i.e.,~$n=37$. These $37$ $\gfourc$-invariants form $\mathcal{V}^4_{37}$. (ii) Even though $\phia^2_{\rs\rda}\phih^2_{\rda\rs}$ and $\phih^2_{\rs\rdg}\phig^2_{\rdg\rs}$ are not $\gfourc$-invariants, their constituent terms $\phia^2_{\rs\rda}$, $\phih^2_{\rda\rs}$, $\phih^2_{\rs\rdg}$ and $\phig^2_{\rdg\rs}$ transform as multiplets under $\gfourc$. (iii) We will show that the map $\phic(c)$ given by  
\begin{align}
\phiac(c)&=\frac{\acute{v}(c)}{\sqrt{3}}\left(\begin{matrix} 1 & -\frac{1}{2} & -\frac{1}{2}\\
0 & \frac{\sqrt{3}}{2} & -\frac{\sqrt{3}}{2}
\end{matrix}\right)^{\!\!\T}\!\!, &\phisac(c)&=\acute{v}_s(c), \notag\\
\phihc(c)&=\hat{v}(c)I/\sqrt{2}, &\phishc(c)&=\hat{v}_s(c),\label{eq:SP}\\
\phigc(c)&=\frac{\grave{v}(c)}{\sqrt{3}}\left(\begin{matrix} 1 &-\frac{1}{2} &  -\frac{1}{2}\\
0 &-\frac{\sqrt{3}}{2} &  \frac{\sqrt{3}}{2}
\end{matrix}\right),&\phisgc(c)&=\grave{v}_s(c),\notag
\end{align}
where $\acute{v}(c)$, $\acute{v}_s(c)$, etc. are the norms of the irreps, is a SP of $\mathcal{V}^4$ for a domain $\mathcal{C}$ that is an open subset of $\mathbb{R}^{39}$. 

Consider the following four group actions under $\gfourc$ on $\phic(c)$ which keep it invariant:
\begin{gather}
T \,\phiac(c)\, (t^2)^\T = \phiac(c),\quad t \, \phigc(c)\, T^\T = \phigc(c),\label{eq:resquasi1}\\
t \,\phihc(c) \,t^\T = \phihc(c),\label{eq:resquasi3}\\
U\,\phiac(c)\,(ut)^\T = \phiac(c),\, ut\,\phigc(c)\,U^\T = \phigc(c),\label{eq:resquasi2}\\ 
ut \,\phihc(c) \,(ut)^\T = \phihc(c).\label{eq:resquasi4}
\end{gather}
The group actions (\ref{eq:resquasi1}), (\ref{eq:resquasi2}) generate a dihedral group, say $D'_6$, and the group actions (\ref{eq:resquasi3}), (\ref{eq:resquasi4}) generate another dihedral group, say $D''_6$. The stabilizer of $\phic(c)$ under $\gfourc$ is nothing but $D'_6\times D''_6$. (\ref{eq:resquasi2}), (\ref{eq:resquasi4}) together form a single group action under $\gfour$. This action generates a cyclic group, say $\cy'_2$, which is the stabilizer of $\phic(c)$ under $\gfour$. We have $D'_6\times D''_6<\gfourc$, $D'_6\times D''_6\nless\gfour$, $\cy'_2<\gfour$ and $\cy'_2<D'_6\times D''_6$ as expected.
 
Orbits with the same conjugacy class of stabilizers are of the same type, and the union of orbits of the same type forms a stratum \cite{MICHEL200111}. In other words, two points belong to the same stratum if and only if their stabilizers are conjugate. The action of the group decomposes the space into strata. An orbit that is isolated in its stratum is called a critical orbit. The points constituting a critical orbit are SPs of every potential irrespective of the coefficients \cite{MICHEL1971, MICHEL200111}. (\ref{eq:resquasi1})-(\ref{eq:resquasi4}) fix all degrees of freedom of $\phic(c)$ except the norms of the irreps. If we quotient out the norms \cite{GOLUBITSKY1988, GAETA2006322}, then in the resulting space, the point corresponding to (\ref{eq:SP}) will be isolated in its stratum under $\gfourc$, and hence it will be a SP of every potential invariant under $\gfourc$. Let $\mathcal{C}_{37}$ be the subspace of $\mathcal{C}$ obtained by equating the coefficients of $\phia^2_{\rs\rda}\phih^2_{\rda\rs}$ and $\phih^2_{\rs\rdg}\phig^2_{\rdg\rs}$ to zero. Let $\phic(c)_{37}$ and $\mathcal{V}^4_{37}$ be the restrictions of $\phic(c)$ (\ref{eq:SP}) and $\mathcal{V}^4$ respectively to $\mathcal{C}_{37}$. We may substitute $\phic(c)_{37}$ in $\mathcal{V}^4_{37}$ and equate the derivatives of $\mathcal{V}^4_{37}$ with respect to the norms to vanish. Any solution to these equations in which all the norms are nonvanishing leads to a SP of the form (\ref{eq:SP}), and the corresponding open subset of $\mathbb{R}^{37}$ is $\mathcal{C}_{37}$. We can show that the terms $\phia^2_{\rs\rda}$, $\phih^2_{\rda\rs}$, $\phih^2_{\rs\rdg}$ and $\phig^2_{\rdg\rs}$ vanish at (\ref{eq:SP}), implying that these terms satisfy the third condition (\ref{eq:assume3}). Therefore, (\ref{eq:SP}) provides $\phic(c)=\phic(c)_{37}$ with $\mathcal{C}=\mathcal{C}_{37} \times\mathbb{R}^2$, and we obtain a set of SPs of $\mathcal{V}^4$ as $g\phic(c) \,\forall g \in \gfourc, c \in \mathcal{C}$.

\vspace{5mm}
\centerline{\bf VII.\hspace{2mm}The framework of the auxiliary group} 
\vspace{1.3mm}
Almost all flavor symmetry groups discussed in the literature so far have been in the form of a direct product of a subgroup of $U(3)$ with Abelian groups, e.g.~$A_4\!\times\!\cy_4\!\times\!\cy_3$. The apparent reason for this assumption is that the fermions exist in three families only. However, it was shown in \cite{1006.0203, 1111.1730, 1211.5143} that by going beyond the $U(3)$-subgroup paradigm we can naturally avoid undesirable coefficients in the potential, similar to the cross-couplings between the two triplets of $A_4$ discussed in \cite{hep-ph/0504165}. References.~\cite{1006.0203, 1111.1730, 1211.5143} used an enlarged flavor group constructed as a semidirect product in which the conventional flavor group [the direct product of a subgroup of $U(3)$ with Abelian groups] appears as the quotient. In more recent works \cite{1901.01205, 1912.02451}, a special case of this construction, in which the semidirect product was replaced with a direct product, was studied. We named this construction the `framework of the auxiliary group' and used the notation $\gf=G_r\!\times\!G_x$ where $\gf$ is the enlarged flavor group, $G_r$ is the conventional flavor group and $G_x$ is the so-called auxiliary group. $G_x$ is defined as the part of the flavor symmetry group under which the fermions remain invariant. Since the fermions transform nontrivially under $G_r$ only, they form an unfaithful representation of $\gf$. The elementary scalar fields, on the other hand, transform nontrivially under both $G_r$ and $G_x$. By taking the tensor products of these elementary fields, we obtain effective fields that transform nontrivially under $G_r$ only so that they can be coupled with the fermions. In this framework, we obtained novel vacuum alignments for the irreps of $G_r$ \cite{1901.01205, 1912.02451}. The current paper provides a firm theoretical foundation for building such alignments. Let us demonstrate this with our example.

We write our flavor symmetry group as the direct product of $G_r=S_4\!\times\!\cy_2$ and $G_x=\acute{D}_6\!\times\!\acute{\cy}_2\!\times\!\grave{D}_6\!\times\!\grave{\cy}_2$. If coupling with fermions requires the scalar fields to transform as $-1$ under $\cy_2$, the lowest order of $\phi$ with which the effective fields can be constructed is cubic. Let us denote such cubic fields with $\xi=(\xi_{\rs}, \xi_{\rd}, \xi_{\rt}, \xi_{\rtp})$ where the subscripts indicate transformation under $S_4$. In the most general form, they are given by 
\begin{align}
\xi_{\rs}&=k_1 \phisa\phish\phisg + k'_1 \text{Tr}[\mathcal{S}],\\
\xi_{\rd}&=k_2\left(\text{Tr}[\mathcal{S}\lambda_8] ,\text{Tr}[\mathcal{S}\lambda_3] \right),\\\xi_{\rt}&=k_3\left(\mathcal{S}_{23}, \mathcal{S}_{31}, \mathcal{S}_{12}\right),\\
\xi_{\rtp}&=k_{\ubar{3}}\left(\mathcal{A}_{23}, \mathcal{A}_{31}, \mathcal{A}_{12}\right),
\end{align}
where $k$'s are arbitrary constants, and $\mathcal{S}$ and $\mathcal{A}$ are the symmetric and the antisymmetric matrices 
\begin{equation}
\mathcal{S} = \frac{1}{2}(\phia \phih \phig + (\phia \phih \phig)^\T), \quad \mathcal{A} = \frac{1}{2}(\phia \phih \phig - (\phia \phih \phig)^\T).
\end{equation}
Let $\xi^{\circ\!}(c)=(\xi_{\rs}^{\circ\!}(c), \xi_{\rd}^{\circ\!}(c), \xi_{\rt}^{\circ\!}(c), \xi_{\rtp}^{\circ\!}(c))$ denote the values of the effective fields at the SP (\ref{eq:SP}). We obtain 
\begin{align}
\xi_{\rs}^{\circ\!}(c)&=k_1\acute{v}_s(c)\hat{v}_s(c)\grave{v}_s(c),\\
\xi_{\rd}^{\circ\!}(c)&=-k_2\acute{v}(c)\hat{v}(c)\grave{v}(c)\frac{1}{\sqrt{6}}(-\frac{1}{2},-\frac{\sqrt{3}}{2}),\\
\xi_{\rt}^{\circ\!}(c)&=k_3\acute{v}(c)\hat{v}(c)\grave{v}(c)\frac{1}{6\sqrt{2}}(2,-1,-1),\\
\xi_{\rtp}^{\circ\!}(c)&=0
\end{align}
The nonvanishing SPs, i.e.,~$\xi_{\rs}^{\circ\!}(c)$, $\xi_{\rd}^{\circ\!}(c)$ and $\xi_{\rt}^{\circ\!}(c)$ contribute towards the construction of the fermion mass matrix. 

Let us focus on $\xi_{\rt}^{\circ\!}(c)\propto (2,-1,-1)$. We have $U\xi_{\rt}^{\circ\!}(c)^\T=\xi_{\rt}^{\circ\!}(c)^\T$. The stabilizer of $\xi_{\rt}^{\circ\!}(c)$ under $G_r = S_4\times Z_2$ is the cyclic group $\cy'_2$ generated by $U$, which is the same as the stabilizer of $\phic(c)$ under $\gfour$ generated by the group action (\ref{eq:resquasi2}), (\ref{eq:resquasi4}). This stabilizer produces the intrinsic constraint $\xi_{\rt_2\!}^{\circ\!}(c)-\xi_{\rt_3\!}^{\circ\!}(c)=0$. We also have a second intrinsic constraint $\xi_{\rt_1\!}^{\circ\!}(c)+2\xi_{\rt_2\!}^{\circ\!}(c)=0$. It is generated as a consequence of $D'_6\times D''_6$, which is the stabilizer of $\phic(c)$ under $\gfourc$. If $\xi_{\rt}$ were an elementary triplet, we would not have been able to obtain this constraint intrinsically. A triplet under $S_4\times Z_2$ represents the symmetry group of a cube. Its normalized space has three critical orbits only \cite{MICHEL200111}: the orbit of $\frac{1}{\sqrt{3}}(1,1,1)$ (vertices), the orbit of $\frac{1}{\sqrt{2}}(1,1,0)$ (edge centers) and the orbit of $(1,0,0)$ (face centers). Therefore, the only way to obtain the SP $\propto(2,-1,-1)$ for an elementary triplet is by constraining the coefficients in the potential. On the other hand, by utilizing the concepts introduced in this work, we obtained the SP $\propto(2,-1,-1)$ for an effective triplet without imposing constraints on the coefficients. In a recent flavor model \cite{1912.02451}, an effective $S_4$ triplet with a similar SP [belonging to the orbit of the SP $\propto(2\sqrt{6},-1,-1)$] was proposed resulting in $\text{TM}_1$ neutrino mixing with $\sin^2 \theta_{13}=\frac{1}{3}\sin^2 \frac{\pi}{12}$. We hope that the present work will form the theoretical basis for such flavor models that aspire to obtain phenomenologically interesting predictions entirely from discrete symmetries without imposing constraints on the coefficients. 

\vspace{5mm}
\centerline{\bf VIII.\hspace{2.0mm}Summary} 
\vspace{1.5mm}
We consider the most general potential of order $p$ invariant under a finite symmetry group $\gfp$. We classify the constraints on a stationary point (SP) of the potential into (a) intrinsic: that originate from the group properties and (b) extrinsic: that are obtained by imposing constraints on the coefficients in the potential. Given a SP, we consider the largest finite group such that its action on the SP produces a set of SPs, independently of the coefficients. We name this group $\gfc$ and discover that it can be larger than $\gfp$. We show that the stabilizer of a SP under not only $\gfp$ but also the larger group $\gfc$ generates intrinsic constraints. We briefly review the framework of the auxiliary group where several elementary fields are coupled together to obtain an effective field. Intrinsic constraints on the SPs of the elementary fields (generated by the stabilizer under $\gfc$) lead to intrinsic constraints on the SP of the effective field. Using an example, we show that these constraints on the SP of the effective field cannot always be obtained intrinsically if the field were elementary. These results have a direct application to flavor models in particle physics. We hope that they find applications in other areas of physics that involve discrete symmetries as well, besides being of general interest in mathematical physics. 

\vspace{5mm}
\centerline{\bf Acknowledgments} 
\vspace{1.5mm}
The author would like to gratefully acknowledge the many stimulating discussions with Sujatha Ramakrishnan. The author would also like to thank the referee of \cite{1901.01205}, whose comments served as an inspiration for this work.


\begin{thebibliography}{27}%
	\makeatletter
	\providecommand \@ifxundefined [1]{%
		\@ifx{#1\undefined}
	}%
	\providecommand \@ifnum [1]{%
		\ifnum #1\expandafter \@firstoftwo
		\else \expandafter \@secondoftwo
		\fi
	}%
	\providecommand \@ifx [1]{%
		\ifx #1\expandafter \@firstoftwo
		\else \expandafter \@secondoftwo
		\fi
	}%
	\providecommand \natexlab [1]{#1}%
	\providecommand \enquote  [1]{``#1''}%
	\providecommand \bibnamefont  [1]{#1}%
	\providecommand \bibfnamefont [1]{#1}%
	\providecommand \citenamefont [1]{#1}%
	\providecommand \href@noop [0]{\@secondoftwo}%
	\providecommand \href [0]{\begingroup \@sanitize@url \@href}%
	\providecommand \@href[1]{\@@startlink{#1}\@@href}%
	\providecommand \@@href[1]{\endgroup#1\@@endlink}%
	\providecommand \@sanitize@url [0]{\catcode `\\12\catcode `\$12\catcode
		`\&12\catcode `\#12\catcode `\^12\catcode `\_12\catcode `\%12\relax}%
	\providecommand \@@startlink[1]{}%
	\providecommand \@@endlink[0]{}%
	\providecommand \url  [0]{\begingroup\@sanitize@url \@url }%
	\providecommand \@url [1]{\endgroup\@href {#1}{\urlprefix }}%
	\providecommand \urlprefix  [0]{URL }%
	\providecommand \Eprint [0]{\href }%
	\providecommand \doibase [0]{http://dx.doi.org/}%
	\providecommand \selectlanguage [0]{\@gobble}%
	\providecommand \bibinfo  [0]{\@secondoftwo}%
	\providecommand \bibfield  [0]{\@secondoftwo}%
	\providecommand \translation [1]{[#1]}%
	\providecommand \BibitemOpen [0]{}%
	\providecommand \bibitemStop [0]{}%
	\providecommand \bibitemNoStop [0]{.\EOS\space}%
	\providecommand \EOS [0]{\spacefactor3000\relax}%
	\providecommand \BibitemShut  [1]{\csname bibitem#1\endcsname}%
	\let\auto@bib@innerbib\@empty
	%</preamble>
	\bibitem [{\citenamefont {King}(2015)}]{1510.02091}%
	\BibitemOpen
	\bibfield  {author} {\bibinfo {author} {\bibfnamefont {S.~F.}\ \bibnamefont
			{King}},\ }\href {\doibase 10.1088/0954-3899/42/12/123001} {\bibfield
		{journal} {\bibinfo  {journal} {J. Phys. G}\ }\textbf {\bibinfo {volume}
			{42}},\ \bibinfo {pages} {123001} (\bibinfo {year} {2015})},\ \Eprint
	{http://arxiv.org/abs/1510.02091} {arXiv:1510.02091 [hep-ph]} \BibitemShut
	{NoStop}%
	\bibitem [{\citenamefont {Petcov}(2018)}]{1711.10806}%
	\BibitemOpen
	\bibfield  {author} {\bibinfo {author} {\bibfnamefont {S.}~\bibnamefont
			{Petcov}},\ }\href {\doibase 10.1140/epjc/s10052-018-6158-5} {\bibfield
		{journal} {\bibinfo  {journal} {Eur. Phys. J. C}\ }\textbf {\bibinfo {volume}
			{78}},\ \bibinfo {pages} {709} (\bibinfo {year} {2018})},\ \Eprint
	{http://arxiv.org/abs/1711.10806} {arXiv:1711.10806 [hep-ph]} \BibitemShut
	{NoStop}%
	\bibitem [{\citenamefont {Hirsch}\ \emph {et~al.}(2010)\citenamefont {Hirsch},
		\citenamefont {Morisi}, \citenamefont {Peinado},\ and\ \citenamefont
		{Valle}}]{1007.0871}%
	\BibitemOpen
	\bibfield  {author} {\bibinfo {author} {\bibfnamefont {M.}~\bibnamefont
			{Hirsch}}, \bibinfo {author} {\bibfnamefont {S.}~\bibnamefont {Morisi}},
		\bibinfo {author} {\bibfnamefont {E.}~\bibnamefont {Peinado}}, \ and\
		\bibinfo {author} {\bibfnamefont {J.}~\bibnamefont {Valle}},\ }\href
	{\doibase 10.1103/PhysRevD.82.116003} {\bibfield  {journal} {\bibinfo
			{journal} {Phys. Rev. D}\ }\textbf {\bibinfo {volume} {82}},\ \bibinfo
		{pages} {116003} (\bibinfo {year} {2010})},\ \Eprint
	{http://arxiv.org/abs/1007.0871} {arXiv:1007.0871 [hep-ph]} \BibitemShut
	{NoStop}%
	\bibitem [{\citenamefont {Haba}\ \emph {et~al.}(2011)\citenamefont {Haba},
		\citenamefont {Kajiyama}, \citenamefont {Matsumoto}, \citenamefont {Okada},\
		and\ \citenamefont {Yoshioka}}]{1008.4777}%
	\BibitemOpen
	\bibfield  {author} {\bibinfo {author} {\bibfnamefont {N.}~\bibnamefont
			{Haba}}, \bibinfo {author} {\bibfnamefont {Y.}~\bibnamefont {Kajiyama}},
		\bibinfo {author} {\bibfnamefont {S.}~\bibnamefont {Matsumoto}}, \bibinfo
		{author} {\bibfnamefont {H.}~\bibnamefont {Okada}}, \ and\ \bibinfo {author}
		{\bibfnamefont {K.}~\bibnamefont {Yoshioka}},\ }\href {\doibase
		10.1016/j.physletb.2010.11.063} {\bibfield  {journal} {\bibinfo  {journal}
			{Phys. Lett. B}\ }\textbf {\bibinfo {volume} {695}},\ \bibinfo {pages} {476}
		(\bibinfo {year} {2011})},\ \Eprint {http://arxiv.org/abs/1008.4777}
	{arXiv:1008.4777 [hep-ph]} \BibitemShut {NoStop}%
	\bibitem [{\citenamefont {Cohn}\ and\ \citenamefont
		{Stewart}(2000)}]{hep-ph/0001333}%
	\BibitemOpen
	\bibfield  {author} {\bibinfo {author} {\bibfnamefont {J.}~\bibnamefont
			{Cohn}}\ and\ \bibinfo {author} {\bibfnamefont {E.}~\bibnamefont {Stewart}},\
	}\href {\doibase 10.1016/S0370-2693(00)00089-7} {\bibfield  {journal}
		{\bibinfo  {journal} {Phys. Lett. B}\ }\textbf {\bibinfo {volume} {475}},\
		\bibinfo {pages} {231} (\bibinfo {year} {2000})},\ \Eprint
	{http://arxiv.org/abs/hep-ph/0001333} {arXiv:hep-ph/0001333} \BibitemShut
	{NoStop}%
	\bibitem [{\citenamefont {Carone}\ \emph {et~al.}(2014)\citenamefont {Carone},
		\citenamefont {Erlich}, \citenamefont {Sher},\ and\ \citenamefont
		{Ramos}}]{1407.6017}%
	\BibitemOpen
	\bibfield  {author} {\bibinfo {author} {\bibfnamefont {C.~D.}\ \bibnamefont
			{Carone}}, \bibinfo {author} {\bibfnamefont {J.}~\bibnamefont {Erlich}},
		\bibinfo {author} {\bibfnamefont {M.}~\bibnamefont {Sher}}, \ and\ \bibinfo
		{author} {\bibfnamefont {R.}~\bibnamefont {Ramos}},\ }\href {\doibase
		10.1103/PhysRevD.90.063521} {\bibfield  {journal} {\bibinfo  {journal} {Phys.
				Rev. D}\ }\textbf {\bibinfo {volume} {90}},\ \bibinfo {pages} {063521}
		(\bibinfo {year} {2014})},\ \Eprint {http://arxiv.org/abs/1407.6017}
	{arXiv:1407.6017 [hep-ph]} \BibitemShut {NoStop}%
	\bibitem [{\citenamefont {Schimmrigk}(2015)}]{1412.8537}%
	\BibitemOpen
	\bibfield  {author} {\bibinfo {author} {\bibfnamefont {R.}~\bibnamefont
			{Schimmrigk}},\ }\href {\doibase 10.1016/j.physletb.2015.06.078} {\bibfield
		{journal} {\bibinfo  {journal} {Phys. Lett. B}\ }\textbf {\bibinfo {volume}
			{748}},\ \bibinfo {pages} {376} (\bibinfo {year} {2015})},\ \Eprint
	{http://arxiv.org/abs/1412.8537} {arXiv:1412.8537 [hep-th]} \BibitemShut
	{NoStop}%
	\bibitem [{\citenamefont {Chigusa}\ and\ \citenamefont
		{Nakayama}(2019)}]{1808.09601}%
	\BibitemOpen
	\bibfield  {author} {\bibinfo {author} {\bibfnamefont {S.}~\bibnamefont
			{Chigusa}}\ and\ \bibinfo {author} {\bibfnamefont {K.}~\bibnamefont
			{Nakayama}},\ }\href {\doibase 10.1016/j.physletb.2018.11.027} {\bibfield
		{journal} {\bibinfo  {journal} {Phys. Lett. B}\ }\textbf {\bibinfo {volume}
			{788}},\ \bibinfo {pages} {249} (\bibinfo {year} {2019})},\ \Eprint
	{http://arxiv.org/abs/1808.09601} {arXiv:1808.09601 [hep-ph]} \BibitemShut
	{NoStop}%
	\bibitem [{\citenamefont {Gaeta}(2006)}]{GAETA2006322}%
	\BibitemOpen
	\bibfield  {author} {\bibinfo {author} {\bibfnamefont {G.}~\bibnamefont
			{Gaeta}},\ }in\ \href {\doibase 10.1016/B0-12-512666-2/00419-3} {\emph
		{\bibinfo {booktitle} {Encyclopedia of Mathematical Physics}}}\ (\bibinfo
	{publisher} {Academic Press},\ \bibinfo {address} {Oxford},\ \bibinfo {year}
	{2006})\ pp.\ \bibinfo {pages} {322--327},\ \Eprint
	{http://arxiv.org/abs/math-ph/0510010} {arXiv:math-ph/0510010} \BibitemShut
	{NoStop}%
	\bibitem [{\citenamefont {Michel}\ and\ \citenamefont
		{Zhilinskii}(2001)}]{MICHEL200111}%
	\BibitemOpen
	\bibfield  {author} {\bibinfo {author} {\bibfnamefont {L.}~\bibnamefont
			{Michel}}\ and\ \bibinfo {author} {\bibfnamefont {B.}~\bibnamefont
			{Zhilinskii}},\ }\href {\doibase 10.1016/S0370-1573(00)00088-0} {\bibfield
		{journal} {\bibinfo  {journal} {Physics Reports}\ }\textbf {\bibinfo {volume}
			{341}},\ \bibinfo {pages} {11} (\bibinfo {year} {2001})},\ \bibinfo {note}
	{symmetry, invariants, topology}\BibitemShut {NoStop}%
         \bibitem [{\citenamefont {Bott}(1982)}]{Bott}%
	\BibitemOpen
	\bibfield  {author} {\bibinfo {author} {\bibfnamefont {R.}~\bibnamefont
			{Bott}},\ }\href {\doibase 10.1090/S0273-0979-1982-15038-8} {\bibfield
		{journal} {\bibinfo  {journal} {Bull. Amer. Math. Soc.}\ }\textbf {\bibinfo {volume}
			{48}},\ \bibinfo {pages} {517--523} (\bibinfo {year} {1982})} \BibitemShut
	{NoStop}%
	\bibitem [{\citenamefont {Altarelli}\ and\ \citenamefont
		{Feruglio}(2005)}]{hep-ph/0504165}%
	\BibitemOpen
	\bibfield  {author} {\bibinfo {author} {\bibfnamefont {G.}~\bibnamefont
			{Altarelli}}\ and\ \bibinfo {author} {\bibfnamefont {F.}~\bibnamefont
			{Feruglio}},\ }\href {\doibase 10.1016/j.nuclphysb.2005.05.005} {\bibfield
		{journal} {\bibinfo  {journal} {Nucl. Phys. B}\ }\textbf {\bibinfo {volume}
			{720}},\ \bibinfo {pages} {64} (\bibinfo {year} {2005})},\ \Eprint
	{http://arxiv.org/abs/hep-ph/0504165} {arXiv:hep-ph/0504165} \BibitemShut
	{NoStop}%
	\bibitem [{\citenamefont {Ma}(2006{\natexlab{a}})}]{hep-ph/0511133}%
	\BibitemOpen
	\bibfield  {author} {\bibinfo {author} {\bibfnamefont {E.}~\bibnamefont
			{Ma}},\ }\href {\doibase 10.1103/PhysRevD.73.057304} {\bibfield  {journal}
		{\bibinfo  {journal} {Phys. Rev. D}\ }\textbf {\bibinfo {volume} {73}},\
		\bibinfo {pages} {057304} (\bibinfo {year} {2006}{\natexlab{a}})},\ \Eprint
	{http://arxiv.org/abs/hep-ph/0511133} {arXiv:hep-ph/0511133} \BibitemShut
	{NoStop}%
	\bibitem [{\citenamefont {Bazzocchi}\ and\ \citenamefont
		{Morisi}(2009)}]{0811.0345}%
	\BibitemOpen
	\bibfield  {author} {\bibinfo {author} {\bibfnamefont {F.}~\bibnamefont
			{Bazzocchi}}\ and\ \bibinfo {author} {\bibfnamefont {S.}~\bibnamefont
			{Morisi}},\ }\href {\doibase 10.1103/PhysRevD.80.096005} {\bibfield
		{journal} {\bibinfo  {journal} {Phys. Rev. D}\ }\textbf {\bibinfo {volume}
			{80}},\ \bibinfo {pages} {096005} (\bibinfo {year} {2009})},\ \Eprint
	{http://arxiv.org/abs/0811.0345} {arXiv:0811.0345 [hep-ph]} \BibitemShut
	{NoStop}%
	\bibitem [{\citenamefont {Chakraborty}\ \emph {et~al.}(2020)\citenamefont
		{Chakraborty}, \citenamefont {Krishnan},\ and\ \citenamefont
		{Ghosal}}]{2003.00506}%
	\BibitemOpen
	\bibfield  {author} {\bibinfo {author} {\bibfnamefont {M.}~\bibnamefont
			{Chakraborty}}, \bibinfo {author} {\bibfnamefont {R.}~\bibnamefont
			{Krishnan}}, \ and\ \bibinfo {author} {\bibfnamefont {A.}~\bibnamefont
			{Ghosal}},\ }\href {\doibase 10.1007/JHEP09(2020)025} {\bibfield  {journal}
		{\bibinfo  {journal} {JHEP}\ }\textbf {\bibinfo {volume} {09}},\ \bibinfo
		{pages} {025} (\bibinfo {year} {2020})},\ \Eprint
	{http://arxiv.org/abs/2003.00506} {arXiv:2003.00506 [hep-ph]} \BibitemShut
	{NoStop}%
	\bibitem [{\citenamefont {King}\ and\ \citenamefont
		{Malinsky}(2007)}]{hep-ph/0610250}%
	\BibitemOpen
	\bibfield  {author} {\bibinfo {author} {\bibfnamefont {S.~F.}\ \bibnamefont
			{King}}\ and\ \bibinfo {author} {\bibfnamefont {M.}~\bibnamefont
			{Malinsky}},\ }\href {\doibase 10.1016/j.physletb.2006.12.006} {\bibfield
		{journal} {\bibinfo  {journal} {Phys. Lett. B}\ }\textbf {\bibinfo {volume}
			{645}},\ \bibinfo {pages} {351} (\bibinfo {year} {2007})},\ \Eprint
	{http://arxiv.org/abs/hep-ph/0610250} {arXiv:hep-ph/0610250} \BibitemShut
	{NoStop}%
	\bibitem [{\citenamefont {King}\ and\ \citenamefont {Luhn}(2010)}]{0912.1344}%
	\BibitemOpen
	\bibfield  {author} {\bibinfo {author} {\bibfnamefont {S.~F.}\ \bibnamefont
			{King}}\ and\ \bibinfo {author} {\bibfnamefont {C.}~\bibnamefont {Luhn}},\
	}\href {\doibase 10.1016/j.nuclphysb.2010.02.019} {\bibfield  {journal}
		{\bibinfo  {journal} {Nucl. Phys. B}\ }\textbf {\bibinfo {volume} {832}},\
		\bibinfo {pages} {414} (\bibinfo {year} {2010})},\ \Eprint
	{http://arxiv.org/abs/0912.1344} {arXiv:0912.1344 [hep-ph]} \BibitemShut
	{NoStop}%
	\bibitem [{\citenamefont {Krishnan}\ \emph {et~al.}(2018)\citenamefont
		{Krishnan}, \citenamefont {Harrison},\ and\ \citenamefont
		{Scott}}]{1801.10197}%
	\BibitemOpen
	\bibfield  {author} {\bibinfo {author} {\bibfnamefont {R.}~\bibnamefont
			{Krishnan}}, \bibinfo {author} {\bibfnamefont {P.}~\bibnamefont {Harrison}},
		\ and\ \bibinfo {author} {\bibfnamefont {W.}~\bibnamefont {Scott}},\ }\href
	{\doibase 10.1140/epjc/s10052-018-5516-7} {\bibfield  {journal} {\bibinfo
			{journal} {Eur. Phys. J. C}\ }\textbf {\bibinfo {volume} {78}},\ \bibinfo
		{pages} {74} (\bibinfo {year} {2018})},\ \Eprint
	{http://arxiv.org/abs/1801.10197} {arXiv:1801.10197 [hep-ph]} \BibitemShut
	{NoStop}%
	\bibitem [{\citenamefont {Pakvasa}\ and\ \citenamefont
		{Sugawara}(1979)}]{Pakvasa:1978tx}%
	\BibitemOpen
	\bibfield  {author} {\bibinfo {author} {\bibfnamefont {S.}~\bibnamefont
			{Pakvasa}}\ and\ \bibinfo {author} {\bibfnamefont {H.}~\bibnamefont
			{Sugawara}},\ }\href {\doibase 10.1016/0370-2693(79)90436-2} {\bibfield
		{journal} {\bibinfo  {journal} {Phys. Lett. B}\ }\textbf {\bibinfo {volume}
			{82}},\ \bibinfo {pages} {105} (\bibinfo {year} {1979})}\BibitemShut
	{NoStop}%
	\bibitem [{\citenamefont {Lee}\ and\ \citenamefont
		{Mohapatra}(1994)}]{hep-ph/9403201}%
	\BibitemOpen
	\bibfield  {author} {\bibinfo {author} {\bibfnamefont {D.-G.}\ \bibnamefont
			{Lee}}\ and\ \bibinfo {author} {\bibfnamefont {R.}~\bibnamefont
			{Mohapatra}},\ }\href {\doibase 10.1016/0370-2693(94)91091-X} {\bibfield
		{journal} {\bibinfo  {journal} {Phys. Lett. B}\ }\textbf {\bibinfo {volume}
			{329}},\ \bibinfo {pages} {463} (\bibinfo {year} {1994})},\ \Eprint
	{http://arxiv.org/abs/hep-ph/9403201} {arXiv:hep-ph/9403201} \BibitemShut
	{NoStop}%
	\bibitem [{\citenamefont {Ma}(2006{\natexlab{b}})}]{hep-ph/0508231}%
	\BibitemOpen
	\bibfield  {author} {\bibinfo {author} {\bibfnamefont {E.}~\bibnamefont
			{Ma}},\ }\href {\doibase 10.1016/j.physletb.2005.10.019} {\bibfield
		{journal} {\bibinfo  {journal} {Phys. Lett. B}\ }\textbf {\bibinfo {volume}
			{632}},\ \bibinfo {pages} {352} (\bibinfo {year} {2006}{\natexlab{b}})},\
	\Eprint {http://arxiv.org/abs/hep-ph/0508231} {arXiv:hep-ph/0508231}
	\BibitemShut {NoStop}%
	\bibitem [{\citenamefont {Michel}(1971)}]{MICHEL1971}%
	\BibitemOpen
	\bibfield  {author} {\bibinfo {author} {\bibfnamefont {L.}~\bibnamefont
			{Michel}},\ }\href{https://gallica.bnf.fr/ark:/12148/bpt6k480300n/f436.texteImage} {\bibfield  {journal} {\bibinfo  {journal} {C. R.
				Acad. Sci. Paris A}\ }\textbf {\bibinfo {volume} {272}},\ \bibinfo {pages}
		{433} (\bibinfo {year} {1971})}\BibitemShut {NoStop}%
	\bibitem [{\citenamefont {Golubitsky}\ \emph {et~al.}(1988)\citenamefont
		{Golubitsky}, \citenamefont {Stewart},\ and\ \citenamefont
		{Schaeffer}}]{GOLUBITSKY1988}%
	\BibitemOpen
	\bibfield  {author} {\bibinfo {author} {\bibfnamefont {M.}~\bibnamefont
			{Golubitsky}}, \bibinfo {author} {\bibfnamefont {I.}~\bibnamefont {Stewart}},
		\ and\ \bibinfo {author} {\bibfnamefont {D.~G.}\ \bibnamefont {Schaeffer}},\
	}in\ \href {\doibase 10.1007/978-1-4612-5034-0} {\emph {\bibinfo {booktitle}
			{Singularities and groups in bifurcation theory}}}\ (\bibinfo  {publisher}
	{Springer},\ \bibinfo {address} {New York},\ \bibinfo {year}
	{1988})\BibitemShut {NoStop}%
	\bibitem [{\citenamefont {Babu}\ and\ \citenamefont
		{Gabriel}(2010)}]{1006.0203}%
	\BibitemOpen
	\bibfield  {author} {\bibinfo {author} {\bibfnamefont {K.}~\bibnamefont
			{Babu}}\ and\ \bibinfo {author} {\bibfnamefont {S.}~\bibnamefont {Gabriel}},\
	}\href {\doibase 10.1103/PhysRevD.82.073014} {\bibfield  {journal} {\bibinfo
			{journal} {Phys. Rev. D}\ }\textbf {\bibinfo {volume} {82}},\ \bibinfo
		{pages} {073014} (\bibinfo {year} {2010})},\ \Eprint
	{http://arxiv.org/abs/1006.0203} {arXiv:1006.0203 [hep-ph]} \BibitemShut
	{NoStop}%
	\bibitem [{\citenamefont {Holthausen}\ and\ \citenamefont
		{Schmidt}(2012)}]{1111.1730}%
	\BibitemOpen
	\bibfield  {author} {\bibinfo {author} {\bibfnamefont {M.}~\bibnamefont
			{Holthausen}}\ and\ \bibinfo {author} {\bibfnamefont {M.~A.}\ \bibnamefont
			{Schmidt}},\ }\href {\doibase 10.1007/JHEP01(2012)126} {\bibfield  {journal}
		{\bibinfo  {journal} {JHEP}\ }\textbf {\bibinfo {volume} {01}},\ \bibinfo
		{pages} {126} (\bibinfo {year} {2012})},\ \Eprint
	{http://arxiv.org/abs/1111.1730} {arXiv:1111.1730 [hep-ph]} \BibitemShut
	{NoStop}%
	\bibitem [{\citenamefont {Holthausen}\ \emph {et~al.}(2013)\citenamefont
		{Holthausen}, \citenamefont {Lindner},\ and\ \citenamefont
		{Schmidt}}]{1211.5143}%
	\BibitemOpen
	\bibfield  {author} {\bibinfo {author} {\bibfnamefont {M.}~\bibnamefont
			{Holthausen}}, \bibinfo {author} {\bibfnamefont {M.}~\bibnamefont {Lindner}},
		\ and\ \bibinfo {author} {\bibfnamefont {M.~A.}\ \bibnamefont {Schmidt}},\
	}\href {\doibase 10.1103/PhysRevD.87.033006} {\bibfield  {journal} {\bibinfo
			{journal} {Phys. Rev. D}\ }\textbf {\bibinfo {volume} {87}},\ \bibinfo
		{pages} {033006} (\bibinfo {year} {2013})},\ \Eprint
	{http://arxiv.org/abs/1211.5143} {arXiv:1211.5143 [hep-ph]} \BibitemShut
	{NoStop}%
	\bibitem [{\citenamefont {Krishnan}(2020)}]{1901.01205}%
	\BibitemOpen
	\bibfield  {author} {\bibinfo {author} {\bibfnamefont {R.}~\bibnamefont
			{Krishnan}},\ }\href {\doibase 10.1103/PhysRevD.101.075004} {\bibfield
		{journal} {\bibinfo  {journal} {Phys. Rev. D}\ }\textbf {\bibinfo {volume}
			{101}},\ \bibinfo {pages} {075004} (\bibinfo {year} {2020})},\ \Eprint
	{http://arxiv.org/abs/1901.01205} {arXiv:1901.01205 [hep-ph]} \BibitemShut
	{NoStop}%
	\bibitem [{\citenamefont {Krishnan}(2019)}]{1912.02451}%
	\BibitemOpen
	\bibfield  {author} {\bibinfo {author} {\bibfnamefont {R.}~\bibnamefont
			{Krishnan}},\ }\href@noop {} {\  (\bibinfo {year} {2019})},\ \Eprint
	{http://arxiv.org/abs/1912.02451} {arXiv:1912.02451 [hep-ph]} \BibitemShut
	{NoStop}%
\end{thebibliography}
\end{document}